\documentclass[preprint,showpacs,preprintnumbers,amsmath,amssymb,showkeys]{revtex4}

\usepackage{graphicx}
\usepackage{dcolumn}
\usepackage{bm}
\usepackage{braket}
\usepackage{enumitem}

\begin{document}

\noindent

\preprint{}

\title{Quantifying quantum coherence via Kirkwood-Dirac quasiprobability}

\author{Agung Budiyono}
\email{agungbymlati@gmail.com}
\author{Hermawan K. Dipojono}
\affiliation{Department of Engineering Physics, Bandung Institute of Technology, Bandung, 40132, Indonesia} 
\affiliation{Research Center for Nanoscience and Nanotechnology, Bandung Institute of Technology, Bandung, 40132, Indonesia}

\date{\today}

\begin{abstract} 

Kirkwood-Dirac (KD) quasiprobability is a quantum analog of phase space probability of classical statistical mechanics, allowing negative or/and nonreal values. It gives an informationally complete representation of a quantum state. Recent works have revealed the important roles played by the KD quasiprobability in the broad fields of quantum science and quantum technology. In the present work, we use the KD quasiprobability to access the quantum coherence in a quantum state. We show that the $l_1$-norm of the imaginary part of the KD quasiprobability over an incoherent reference basis and a second basis, maximized over all possible choices of the latter, can be used to quantify quantum coherence, satisfying certain desirable properties. It is upper bounded by the quantum uncertainty, i.e., the quantum standard deviation, of the incoherent basis in the state. It gives a lower bound to the $l_1$-norm quantum coherence, and for a single qubit, they are identical. We discuss the measurement of the KD coherence based on the measurement of the KD quasiprobability and an optimization procedure in hybrid quantum-classical schemes, and suggest statistical interpretations. We also discuss its relevance in the physics of linear response regime. 

\end{abstract} 

\maketitle       

\section{Introduction}   

Quantum coherence is one of the defining features of quantum mechanics, manifesting the superposition principle. It underlies the nonclassical features of quantum phenomena. Recently, quantum coherence has also been recognized as one of the key ingredients for various schemes of quantum technologies \cite{Streltsov review,Marvian - Spekkens speakable and unspeakable coherence}. In the last decade, the success of the resource theoretical framework to study diverse nonclassical features of quantum systems by regarding them as constituting resources for some operational tasks \cite{Horodecki resource theory}, has led many researchers to apply the framework to rigorously characterize quantum coherence \cite{Marvian - Spekkens speakable and unspeakable coherence,Aberg quantifying of superposition,Levi quantum coherence measure,Vogel quantum coherence measure,Baumgratz quantum coherence measure,Girolami quantum coherence measure,Winter operational resource theory of coherence,Streltsov review,Marvian coherence measure,Chitambar physically consistent resource theory of coherence,Napoli robustness of coherence,Yu alternative resource theory of coherence,Yuan quantum coherence intrinsic randomness}. In this approach, one defines coherence as an aspect which cannot be created by different classes of incoherence preserving quantum operations. However, while mathematically well-defined, the physical interpretation of these formal operations are not entirely clear \cite{Marvian - Spekkens speakable and unspeakable coherence,Marvian coherence measure,Chitambar physically consistent resource theory of coherence}. Moreover, the resulting coherence quantifiers do not have transparent interpretation in terms of direct laboratory operations. 

On the other hand, recently, there has been a revival of interest in the Kirkwood-Dirac (KD) quasiprobability, an informationally complete representation of a quantum state \cite{Kirkwood quasiprobability,Dirac quasiprobability,Barut KD quasiprobability,Chaturvedi KD distribution}. KD quasiprobability returns correct marginal probabilities, but it may take negative or/and nonreal values. Such negativity or nonreality, a.k.a. KD nonclassicality, indicates nonclassicality stronger than noncommutativity \cite{Drori nonclassicality tighter and noncommutativity,deBievre nonclassicality in KD distribution}, and is suggested as the origin of quantum advantage in certain quantum metrology \cite{Arvidsson-Shukur quantum advantage in postselected metrology} and quantum heat engine \cite{Lostaglio contextuality in quantum linear response}. KD quasiprobability appears naturally in different forms of quantum fluctuations, and KD nonclassicality has been argued to signify genuine quantum behaviour of the underlying physical processes \cite{Lostaglio KD quasiprobability and quantum fluctuation}. It has been used to characterize work distribution to extend thermodynamics fluctuation theorem in quantum regime \cite{Allahverdyan TMH as quasiprobability distribution of work,Lostaglio TMH quasiprobability fluctuation theorem contextuality}, as a witness of information scrambling in many body systems \cite{Alonso KD quasiprobability witnesses quantum scrambling,Halpern quasiprobability and information scrambling}, and as proofs of contextuality \cite{Pusey negative TMH quasiprobability and contextuality,Kunjwal KD quasiprobability and contextuality}. It is therefore instructive to ask: how does coherence in a quantum state is encoded in the associated KD quasiprobability representation. The answer to this question might also offer useful insight into the roles of quantum coherence in physical situations listed above where KD nonclassicality is crucial. 

In the present work, we propose a characterization and quantification of quantum coherence based on KD quasiprobability. First, given a quantum state and an incoherent reference basis, we identify a quantity, referred to as KD coherence, that is given by the $l_1$-norm of the imaginary part of the KD quasiprobability defined over a reference basis and a second basis, and maximized over all possible choices of the latter. It formalizes the intuition that coherence should reflect the noncommutativity between the state and the incoherent basis, and we show that it satisfies certain desirable properties for a quantifier of quantum coherence. It is upper bounded by the total sum of the quantum standard deviation, thus the quantum uncertainty, of the incoherent basis in the state. KD coherence gives a lower bound to the $l_1$-norm coherence, and for an arbitrary state of a single qubit, they give the same value. We discuss the observation of the KD coherence via a couple of methods for the reconstruction of KD quasiprobability, combined with an optimization procedure in hybrid quantum-classical schemes. These suggest statistical interpretation of the KD coherence as the maximal disturbance induced by the measurement of, or the maximal mean absolute error in the estimation of the incoherent basis. We also give a short discussion on the relevance of the KD coherence to characterize linear response function.  

\section{Quantum coherence and Kirkwood-Dirac quasiprobability} 

\subsection{Quantum coherence}

Consider a quantum system with the Hilbert space of finite dimension $d$, and choose an orthonormal basis $\{\ket{a}\}$, $\sum_a\Pi_a=\mathbb{I}$, where $\Pi_a:=\ket{a}\bra{a}$ is a projector, assumed, for simplicity, to be one dimensional (rank-one projector). Such a basis decomposes the $d$ dimensional Hilbert space into the direct sum of the one-dimensional $d$ subspaces, each is spanned by $\ket{a}$. A quantum state represented by the density operator $\varrho$ on the Hilbert space is said to be incoherent with respect to the reference basis $\{\ket{a}\}$ (or, relative to the Hilbert space decomposition into the associated subspaces) if it can be expressed as 
\begin{eqnarray}
\varrho=\sum_a p_a\ket{a}\bra{a}, 
\label{incoherent state}
\label{eqnarray}
\end{eqnarray}
$p_a=\braket{a|\varrho|a}$, $\sum_ap_a=1$. Namely, it is a classical statistical mixture of the elements of the reference basis. Hence, the density operator is diagonal with respect to the reference basis so that they are commuting, i.e., $[\Pi_a,\varrho]=0$, for all $a$. Any state that cannot be so expressed is coherent with respect to the basis $\{\ket{a}\}$. In this sense, $\{\ket{a}\}$ is referred to as the incoherent reference basis. The choice of the incoherent reference basis depends on the physical problem and/or the physical system under investigation. 

A mathematically rigorous information theoretical framework to characterize coherence by regarding it as a resource is attracting a lot of attention recently \cite{Marvian - Spekkens speakable and unspeakable coherence,Aberg quantifying of superposition,Levi quantum coherence measure,Vogel quantum coherence measure,Baumgratz quantum coherence measure,Girolami quantum coherence measure,Winter operational resource theory of coherence,Streltsov review,Marvian coherence measure,Chitambar physically consistent resource theory of coherence,Napoli robustness of coherence,Yu alternative resource theory of coherence,Yuan quantum coherence intrinsic randomness}. In this resource theoretical framework \cite{Horodecki resource theory}, quantum states and operations are divided into those that are free and those whose preparation and implementation bear some cost. For example, in the resource theory of entanglement, the free operations are identified by the local operation and classical communication (LOCC) so that the free states are given by unentangled (separable) states \cite{Vedral resource theory of entanglement,Horodecki resource theory of entanglement}. Such a division intuitively reflects the operational restriction in experimental scenario involving distant parties. In this framework, entangled states are thus seen as states with a resource whose provision may be used to overcome the restriction. Analogously, in the resource theory of coherence, the incoherent quantum states of Eq. (\ref{incoherent state}) are assumed to be free, and the free operations are given by several different classes of incoherence preserving quantum operations \cite{Marvian - Spekkens speakable and unspeakable coherence,Streltsov review}. Quantum coherence is therefore naturally defined as the resource that cannot be created by these operations. This approach has led to the construction of various important coherence quantifiers. However, unlike LOCC, it is difficult to give a clear interpretation to the incoherence preserving operations alluded to above in terms of operational restriction in laboratory \cite{Marvian - Spekkens speakable and unspeakable coherence,Chitambar physically consistent resource theory of coherence,Marvian coherence measure}. Moreover, most of the resulting coherence quantifiers cannot be interpreted in terms of direct laboratory operations \cite{Napoli robustness of coherence}. 

For later reference, let us summarize the $l_1$-norm coherence arising in the above resource-theoretic approach \cite{Baumgratz quantum coherence measure}. Consider an arbitrary quantum state $\varrho=\sum_{a,a'}\varrho_{aa'}\ket{a}\bra{a'}$, $\varrho_{aa'}=\braket{a|\varrho|a'}$, where $\{\ket{a}\}$ is the incoherent basis. The $l_1$-norm quantum coherence in $\varrho$ relative to the incoherent basis $\{\ket{a}\}$ is then defined as: $C_{l_1}[\varrho;\{\Pi_a\}]:=\min_{\tau\in\mathcal{I}\{\ket{a}\}}\|\varrho-\tau\|_{l_1}=\sum_{a\neq a'}|\varrho_{aa'}|$, where $\mathcal{I}\{\ket{a}\}$ is the set of all incoherent states relative to the reference basis $\{\ket{a}\}$, and $\|\cdot\|_{l_1}$ is the $l_1$ matrix norm. Hence, it is given by the sum of the absolute value of the off-diagonal terms of the density matrix, directly capturing the intuition that coherence must quantify the interference between the elements of the reference basis. Remarkably, for a single qubit, various different coherence quantifiers are equal to, or can be written as a simple function of, the $l_1$-norm coherence \cite{Streltsov review}. The $l_1$-norm coherence can be used to quantify the wave aspect in the wave-particle complementarity relations \cite{Bera quantum coherence and 2 paths wave-particle duality,Prillwitz coherence multipath interference,Bagan quantum coherence and multipaths wave-particle duality,Biswas interferometric visibility and coherence,Paul measurement of coherence in multislit interference,Qureshi measurement of coherence in multipath interference}. It also has proven to be useful in studying speedup in quantum computation \cite{Deutch algorithm and coherence,Hillery coherence in decision problems,Shi Grover algorithm and coherence,Anand coherence in discrete Grover search,Matera coherence in quantum algorithm,Ma coherence in quantum algorithm}.   

\subsection{Kirkwood-Dirac quasiprobability}

There is an informationally equivalent representation of the quantum state based on quasiprobability. Quasiprobability is the quantum analog of phase space probability distribution for classical statistical mechanics \cite{Lee quasiprobability review}. Due to the quantum noncommutativity (incompatibility), quasiprobability necessarily does not satisfy all the Kolmogorov's axioms for conventional probability \cite{Lostaglio KD quasiprobability and quantum fluctuation}. For example, the Wigner function, the most well-known quasiprobability, may take negative values. There are infinitely many quasiprobability representation arising from the ambiguity of the ordering of operators. Here, for system with finite dimensional Hilbert space, and for the reason that will be clarified later, we shall use the representation of quantum state in terms of a specific quasiprobability called as Kirkwood-Dirac (KD) quasiprobability \cite{Kirkwood quasiprobability,Dirac quasiprobability,Barut KD quasiprobability,Chaturvedi KD distribution} to access the coherence in the quantum state. 
 
Given a quantum state $\varrho$ acting on a Hilbert space with dimension $d$, and two bases $\{\ket{a}\}$ and $\{\ket{b}\}$ of the Hilbert space, the KD quasiprobability is defined as 
\begin{eqnarray}
{\rm Pr}_{\rm KD}(a,b|\varrho):={\rm Tr}\{\Pi_b\Pi_a\varrho\}=\braket{b|\Pi_a\varrho|b}.
\label{Kirkwood-Dirac quasiprobability}
\end{eqnarray}
The KD quasiprobability gives correct marginal probabilities, i.e., $\sum_a{\rm Pr}_{\rm KD}(a,b|\varrho)={\rm Tr}\{{\Pi_b\varrho}\}$, $\sum_b{\rm Pr}_{\rm KD}(a,b|\varrho)={\rm Tr}\{{\Pi_a\varrho}\}$, and thus normalized $\sum_{a,b}{\rm Pr}_{\rm KD}(a,b|\varrho)=1$, but, it may assume negative or/and non-real values capturing nonclassicality tighter than noncommutativity \cite{Drori nonclassicality tighter and noncommutativity,deBievre nonclassicality in KD distribution}. The real part is known as Terletsky-Margenau-Hill quasiprobability \cite{Terletsky TMH quasiprobability,Margenau TMH quasiprobability}. Given the KD quasiprobability ${\rm Pr}_{\rm KD}(a,b|\varrho)$, the density matrix $\varrho$ can be recovered as, assuming $\braket{a|b}\neq 0$ for all $(a,b)$, $\sum_{a,b}{\rm Pr}_{\rm KD}(a,b|\varrho)\frac{\ket{a}\bra{b}}{\braket{b|a}}=\sum_{a,b}\braket{a|\varrho|b}\ket{a}\bra{b}=\varrho$, hence, they are informationally equivalent. Choosing a pair of bases so that $\braket{a|b}=\frac{1}{\sqrt{d}}e^{i2\pi ab/d}$, the density matrix in the basis $\{\ket{a}\}$ is thus obtained by Fourier transforming the KD quasiprobability as $\braket{a|\varrho|a'}=\sum_{b=0}^{d-1}{\rm Pr}_{\rm KD}(a,b|\varrho) e^{i\frac{2\pi}{d}(a-a')b}$ \cite{Chaturvedi KD distribution}. One of the advantages of using KD quasiprobability representation is that one may use the negativity or/and the nonreality of the KD quasiprobability, i.e., the KD nonclassicality, to access genuine nonclassical behaviour of a quantum system, by showing that it violates some classical bound derived based on conventional real and non-negative probability. Indeed, as listed in the Introduction, the KD nonclassicality is playing significant roles in the study of quantum information \cite{Arvidsson-Shukur quantum advantage in postselected metrology,Lostaglio contextuality in quantum linear response}, quantum fluctuation \cite{Lostaglio KD quasiprobability and quantum fluctuation,Alonso KD quasiprobability witnesses quantum scrambling,Halpern quasiprobability and information scrambling}, quantum thermodynamics \cite{Allahverdyan TMH as quasiprobability distribution of work,Lostaglio TMH quasiprobability fluctuation theorem contextuality}, and quantum foundation \cite{Pusey negative TMH quasiprobability and contextuality,Kunjwal KD quasiprobability and contextuality}. 

\section{Quantum coherence from the imaginary part of the KD quasiprobability}

Since KD quasiprobability is an informationally complete representation of the quantum state, it is natural to ask how the KD quasiprobability representation encodes the quantum coherence in the quantum state relative to a given incoherent basis. Note that the KD quasiprobability is defined in terms of two bases, while quantum coherence is defined relative to a single incoherent basis. To pursue this question, we observe first a simple fact that for an arbitrary quantum state $\varrho$ and a basis $\{\ket{a}\}$, the imaginary part of the corresponding KD quasiprobability captures the commutation relation between the state and the basis, i.e.,  
\begin{eqnarray}
{\rm Im}\{{\rm Pr}_{\rm KD}(a,b|\varrho)\}&=&{\rm Im}\{\braket{b|\Pi_a\varrho|b}\}=\frac{1}{2i}\braket{b|[\Pi_a,\varrho]|b}\nonumber\\
&=&\sum_{a'\neq a}{\rm Im}\{\varrho_{aa'}\braket{b|a}\braket{a'|b}\}. 
\label{incompatibility and coherence}
\end{eqnarray}
It is also clear from the second line that, choosing a second basis $\{\ket{b}\}$ such that $\braket{b|a}\braket{a'|b}\neq 0$ for some pairs of $(a,a')$, $a\neq a'$, ${\rm Im}\{{\rm Pr}_{\rm KD}(a,b|\varrho)\}\neq 0$ implies that not all of the off-diagonal terms of the density matrix are vanishing, indicating the presence of coherence in $\varrho$ with respect to the incoherent reference basis $\{\ket{a}\}$. 

We wish to devise a simple quantity from the imaginary part of the KD quasiprobability, which can faithfully detect the quantum coherence, and possesses certain properties expected for a coherence quantifier. To this end, given a general quantum state $\varrho$ and an incoherent reference basis $\{\ket{a}\}$, let us define the following quantity which maps the quantum state to a nonnegative quantity:
\begin{eqnarray}
C_{\rm KD}[\varrho;\{\Pi_a\}]&:=&\max_{\{\ket{b}\}}\sum_a\sum_b\big|{\rm Im}\{{\rm Pr}_{\rm KD}(a,b|\varrho)\}\big|\nonumber\\
&=&\max_{\{\ket{b}\}}\sum_a\sum_b\big|{\rm Im}\{\braket{b|\Pi_a\varrho|b}\}\big|\nonumber\\
&=&\max_{\{\ket{b}\}}\sum_a\sum_b\frac{1}{2}\big|\braket{b|[\Pi_a,\varrho]|b}\big|,
\label{KD coherence}
\end{eqnarray}
where $\{\ket{b}\}$ is another basis of the Hilbert space. We thus take the $l_1$-norm of ${\rm Pr}_{\rm KD}(a,b|\varrho)$ and maximize over all possible choices of the second basis $\{\ket{b}\}$. The maximization seeks for the largest incompatibility between the quantum state $\varrho$ and the incoherent basis $\{\ket{a}\}$, with respect to the the second basis $\{\ket{b}\}$, under the $l_1$-norm. Next, suppose we wish to quantify the coherence of a composite of $N$ subsystems with respect to an incoherent product basis, i.e., $\{\ket{a}\}=\{\ket{a_1}\otimes\cdots\otimes\ket{a_N}\}:=\{\ket{a_1,\dots,a_N}\}$, where $\ket{a_i}$ is the first basis for subsystem $i$. Then, we assume that the second basis is also a product, i.e., $\{\ket{b}\}=\{\ket{b_1,\dots,b_N}\}$ where $\{\ket{b_i}\}$ is the second basis for subsystem $i$. 

We show that $C_{\rm KD}[\varrho;\{\Pi_a\}]$, here on referred to as KD coherence, satisfies certain desirable properties for a quantifier of quantum coherence, as follows:
\begin{enumerate}[label=(\roman*)]
\item Faithful, i.e., $C_{\rm KD}[\varrho;\{\Pi_a\}]=0$ if and only if the quantum state $\varrho$ is incoherent with respect to the basis $\{\ket{a}\}$. 
\item Convex, i.e., $C_{\rm KD}[\sum_kp_k\varrho_k;\{\Pi_a\}]\le\sum_kp_kC_{\rm KD}[\varrho_k;\{\Pi_a\}]$, where $\{p_k\}$ are probabilities: $0\le p_k\le 1$, $\sum_kp_k=1$.
\item Unitarily covariant: $C_{\rm KD}[U\varrho U^{\dagger};\{ U\Pi_a U^{\dagger}\}]=C_{\rm KD}[\varrho;\{\Pi_a\}]$.
\item Invariant under unitary transformations which commute with a Hermitian observable whose eigenvectors are given by the incoherent basis: $C_{\rm KD}[ U_A\varrho U_A^{\dagger};\{\Pi_a\}]=C_{\rm KD}[\varrho;\{\Pi_a\}]$, where $[U_A,A]=0$, $A=\sum_aa\Pi_a$, $a\in\mathbb{R}$.  
\item Invariant under unitary transformation which permutes the index of the elements in the incoherent basis: $C_{\rm KD}[ U_{\rm p}\varrho U_{\rm p}^{\dagger};\{\Pi_a\}]=C_{\rm KD}[\varrho;\{\Pi_a\}]$, where $ U_{\rm p}\ket{a}=e^{i\theta_a}\ket{\mu(a)}$, $\mu(a)$ is a permutation of index in the basis, and $\theta_a\in\mathbb{R}$.
\item Nonincreasing under partial trace: $C_{\rm KD}[\varrho_{12};\{\Pi_{a_1}\otimes\mathbb{I}_2\}]\ge C_{\rm KD}[\varrho_1;\{\Pi_{a_1}\}]$, where $\varrho_{12}$ is the quantum state of the composite of subsystem $1$ and $2$, $\varrho_1={\rm Tr}_2\{\varrho_{12}\}$ is the quantum state of subsystem 1, $\{\ket{a_1}\}$ is the incoherent basis of subsystem 1, and $\mathbb{I}_2$ is the identity operator of subsystem 2. 
\item Nonincreasing under decoherence operation, i.e., $C_{\rm KD}[\varrho;\{\Pi_a\}]\ge C_{\rm KD}[\varrho';\{\Pi_a\}]$, where $\varrho'=p\varrho+(1-p)\mathcal{D}(\varrho;\{\Pi_a\})$, $0\le p\le 1$, and $\mathcal{D}(\varrho;\{\Pi_a\}):=\sum_a\Pi_a\varrho\Pi_a$ is the dephasing operation which removes the off-diagonal terms of $\varrho$ in the basis $\{\ket{a}\}$. 
\end{enumerate}

Let us sketch and discuss the proofs of the above properties. 

To establish property (i) of faithfulness, first note that if $\varrho$ is an incoherent state so that $[\Pi_a,\varrho]=0$ for all $a$, we have $C_{\rm KD}[\varrho;\{\Pi_a\}]=0$ by definition. Conversely, let us suppose that $C_{\rm KD}[\varrho;\{\Pi_a\}]=0$. Then, from the definition, we must have ${\rm Im}\{{\rm Pr}_{\rm KD}(a,b|\varrho)\}=\braket{b|[\Pi_a,\varrho]|b}/2i=0$ for all $a$ and $b$. This can only be true for all possible choices of $\{\ket{b}\}$ if $[\Pi_a,\varrho]=0$ for all $a$. This means that $\{\Pi_a\}$ is the eigenprojectors for $\varrho$, so that $\varrho$ must be expressible as in Eq. (\ref{incoherent state}), i.e., it is incoherent relative to the reference basis $\{\ket{a}\}$. 

Next, property (ii) of convexity shows that classical mixing $\varrho=\sum_kp_k\varrho_k$ does not increase KD coherence, suggesting that it quantifies a genuine quantum information. This is a trivial implication of the triangle inequality for the $l_1$-norm and the fact that $p_k\ge 0$, i.e., $C_{\rm KD}[\sum_kp_k\varrho_k;\{\Pi_a\}]=\max_{\{\ket{b}\}}\sum_a\sum_b\big|{\rm Im}\{\braket{b|\Pi_a\sum_kp_k\varrho_k|b}\}\big|\le\sum_kp_k\max_{\{\ket{b}\}}\sum_a\sum_b\big|{\rm Im}\{\braket{b|\Pi_a\varrho_k|b}\}\big|=\sum_kp_kC_{\rm KD}[\varrho_k;\{\Pi_a\}]$. 

The property (iii) of unitarily covariant can be directly established from the definition, i.e., 
\begin{eqnarray}
&&C_{\rm KD}[ U\varrho U^{\dagger};\{ U\Pi_a U^{\dagger}\}]\nonumber\\
&=&\max_{\{\ket{b}\}}\sum_a\sum_b\big|{\rm Im}\{\braket{b| U\Pi_a U^{\dagger} U\varrho U^{\dagger}|b}\}\big|\nonumber\\
&=&\max_{\{\ket{b'}\}}\sum_a\sum_{b'}\big|{\rm Im}\{\braket{b'|\Pi_a\varrho|b'}\}\big|\nonumber\\
&=&C_{\rm KD}[\varrho;\{\Pi_a\}], 
\label{proof of the unitary covariant property}
\end{eqnarray}
where we have taken into account the fact that unitary operator $U$ leads to transformation between bases $\{\ket{b'}\}=\{U^{\dagger}\ket{b}\}$ of the same Hilbert space, so that $\max_{\{\ket{b'}\}}(\cdot)=\max_{\{\ket{b}\}}(\cdot)$. This property captures the intuition that simultaneously unitarily rotating both the incoherent basis and the quantum state in the Hilbert space should give the same value of coherence. 

To establish property (iv), we first note that for any unitary operator $U_A$ which commutes with $A=\sum_a a\ket{a}\bra{a}$, we have $U_A\ket{a}=e^{i\theta_a}\ket{a}$, $\theta_a\in\mathbb{R}$, so that 
\begin{eqnarray}
&&C_{\rm KD}[ U_A\varrho U_A^{\dagger};\{\Pi_a\}]\nonumber\\
&=&\max_{\{\ket{b}\}}\sum_a\sum_b\big|{\rm Im}\{\braket{b|U_A U_A^{\dagger}\Pi_a U_A\varrho U_A^{\dagger}|b}\}\big|\nonumber\\
&=&\max_{\{\ket{b'}\}}\sum_a\sum_{b'}\big|{\rm Im}\{\braket{b'|\Pi_a\varrho|b'}\}\big|\nonumber\\
&=&C_{\rm KD}[\varrho;\{\Pi_a\}],
\end{eqnarray}
where in the second line we have inserted the identity $U_AU_A^{\dagger}=\mathbb{I}$, and in the third line we defined $\{\ket{b'}\}=\{U_A^{\dagger}\ket{b}\}$ and used the fact that $\max_{\{\ket{b'}\}}(\cdot)=\max_{\{\ket{b}\}}(\cdot)$. We note that such unitaries which commute with $A$ is covariant under the translation $U=e^{-iA\theta}$ generated by $A$ (taking $\hbar=1$), in the sense that its implementation followed by the translation yields the same result when the order of the operations is reversed: $e^{-iA\theta}U_A\varrho U_A^{\dagger}e^{iA\theta}=U_Ae^{-iA\theta}\varrho e^{iA\theta}U_A^{\dagger}$ \cite{Marvian - Spekkens speakable and unspeakable coherence}. 

Next, consider a unitary operator which permutes the elements of the incoherent basis, i.e., $U_{\rm p}=\sum_ae^{i\theta_a}\ket{\mu(a)}\bra{a}$, where $\mu(a)$ is an index permutation. Such a permutation of index in the reference basis should not change the coherence relative to the basis as claimed by property (v). To see this, first we have $\{U_{\rm p}\Pi_a U_{\rm p}^{\dagger}\}=\{\Pi_{\mu(a)}\}=\{\Pi_a\}$. Noting this, we may proceed as 
\begin{eqnarray}
&&C_{\rm KD}[ U_{\rm p}\varrho U_{\rm p}^{\dagger};\{\Pi_a\}]\nonumber\\
&=&\max_{\{\ket{b}\}}\sum_a\sum_b\big|{\rm Im}\{\braket{b|U_{\rm p} U_{\rm p}^{\dagger}\Pi_a U_{\rm p}\varrho U_{\rm p}^{\dagger}|b}\}\big|\nonumber\\
&=&\max_{\{\ket{b'}\}}\sum_a\sum_{b'}\big|{\rm Im}\{\braket{b'|\Pi_{\mu(a)}\varrho|b'}\}\big|\nonumber\\
&=&\max_{\{\ket{b'}\}}\sum_a\sum_{b'}\big|{\rm Im}\{\braket{b'|\Pi_a\varrho|b'}\}\big|\nonumber\\
&=&C_{\rm KD}[\varrho;\{\Pi_a\}],
\end{eqnarray}
where we have inserted $U_{\rm p}U_{\rm p}^{\dagger}=\mathbb{I}$ and defined $\{\ket{b'}\}=\{U_{\rm p}^{\dagger}\ket{b}\}$, and in the fourth line we have relabelled the sum over $a$. We note that the set of $U_{\rm p}$ for a given reference basis comprises all the incoherence preserving unitaries, which is equivalent to the set of dephasing covariant unitaries \cite{Marvian - Spekkens speakable and unspeakable coherence}, i.e., those unitaries whose operation followed by the dephasing operation $\mathcal{D}(\varrho;\{\Pi_a\})$ yield the same effect when the order of the operations is reversed. 

Property (vi) captures the intuition that if two subsystems are correlated, ignoring one of them should not increase the coherence of the other. This can be shown as
\begin{eqnarray}
&&C_{\rm KD}[\varrho_{12};\{\Pi_{a_1}\otimes\mathbb{I}_2\}]\nonumber\\
&:=&\max_{\{\ket{b_1,b_2}\}}\sum_{a_1}\sum_{b_1,b_2}\big|{\rm Im}\{\sum_{a_2}{\rm Pr}_{\rm KD}(a_1,a_2,b_1,b_2|\varrho_{12})\}\big|\nonumber\\
&=&\max_{\{\ket{b_1,b_2}\}}\sum_{a_1}\sum_{b_1,b_2}\big|{\rm Im}\{\braket{b_1,b_2|(\Pi_{a_1}\otimes\mathbb{I}_2)\varrho_{12}|b_1,b_2}\}\big|\nonumber\\
&\ge&\max_{\{\ket{b_1,b_2}\}}\sum_{a_1}\sum_{b_1}\big|{\rm Im}\big\{\sum_{b_2}\braket{b_1,b_2|(\Pi_{a_1}\otimes\mathbb{I}_2)\varrho_{12}|b_1,b_2}\big\}\big|\nonumber\\
&=&\max_{\{\ket{b_1}\}}\sum_{a_1}\sum_{b_1}\big|{\rm Im}\{\braket{b_1|\Pi_{a_1}\varrho_1|b_1}\}\big|\nonumber\\
&=& C_{\rm KD}[\varrho_1;\{\Pi_{a_1}\}],
\end{eqnarray}
where $\varrho_1=\sum_{b_2}\braket{b_2|\varrho_{12}|b_2}={\rm Tr}_2\{\varrho_{12}\}$. One can see that the equality is obtained when there is no quantum and classical correlation in the quantum state, i.e., $\varrho_{12}=\varrho_1\otimes\varrho_2$, by virtue of the fact that $\braket{b_2|\varrho_2|b_2}$ is real and positive for all $b_2$, and $\sum_{b_2}\braket{b_2|\varrho_2|b_2}=1$. 

Finally, property (vii) can be shown as follows:
\begin{eqnarray}
&&C_{\rm KD}[p\varrho+(1-p)\mathcal{D}(\varrho;\{\Pi_{a'}\});\{\Pi_a\}]\nonumber\\
&=&pC_{\rm KD}[\varrho;\{\Pi_a\}]\le C_{\rm KD}[\varrho;\{\Pi_a\}]],
\end{eqnarray}
where, we have used the fact that $[\mathcal{D}(\varrho;\{\Pi_{a'}\}),\Pi_a]=0$ for all $a$ and $p\ge 0$ to get the equality in the second line.  

Let us discuss a few implications of the above definition of KD coherence. First, it is clear from the definition that the maximum KD coherence in a quantum state relative to all possible incoherent bases is obtained as the maximum of the $l_1$-norm of the imaginary part of the associated KD quasiprobability defined over all possible pair of bases, i.e., $\max_{\{\ket{a}\}}C_{\rm KD}[\varrho;\{\Pi_a\}]=\max_{\{\ket{a}\}}\max_{\{\ket{b}\}}\sum_a\sum_b\big|{\rm Im}\{{\rm Pr}_{\rm KD}(a,b|\varrho)\}\big|$. Or, equivalently, the maximum of the $l_1$-norm of the imaginary part of the associated KD quasiprobability over all pair of the defining bases encodes the maximum coherence in the state relative to all incoherent bases.     

Next, since KD coherence is defined as the maximal incompatibility between the state and the incoherent basis, it is natural to expect that it somewhat captures the genuine quantum uncertainty of the basis in the quantum state. It is therefore instructive to compare KD coherence relative to a basis with the quantum variance of the basis. Note that quantum variance quantifies the total quantum uncertainty which also includes the uncertainty arising from classical mixing. We show that KD coherence $C_{\rm KD}[\varrho;\{\Pi_a\}]$ is always lower than or equal to the total sum of the square root of the quantum variance (i.e., quantum standard deviation) of the basis $\{\Pi_a\}$ in the state $\varrho$. To see this, we first have, from Eq. (\ref{KD coherence}), 
\begin{eqnarray}
&&C_{\rm KD}[\varrho ;\{\Pi_a\}]\nonumber\\
&=&\max_{\{\ket{b}\}}\sum_a\sum_b\Big|{\rm Im}\Big\{\frac{{\rm Tr}\{\Pi_b\Pi_a\varrho\}}{{\rm Tr}\{\Pi_b\varrho\}}\Big\}\Big|{\rm Tr}\{\Pi_b\varrho\}\nonumber\\
&\le&\sum_a\Big[\sum_{b_*}\Big(\Big|\frac{{\rm Tr}\{\Pi_{b_*}\Pi_a\varrho\}}{{\rm Tr}\{\Pi_{b_*}\varrho\}}\Big|^2-{\rm Re}\Big\{\frac{{\rm Tr}\{\Pi_{b_*}\Pi_a\varrho\}}{{\rm Tr}\{\Pi_{b_*}\varrho\}}\Big\}^2\Big){\rm Tr}\{\Pi_{b_*}\varrho\}\Big]^{1/2}\nonumber\\
&\le&\sum_a\Big[\sum_{b_*}\frac{({\rm Tr}\{\Pi_{b_*}\Pi_a\varrho \})^2}{{\rm Tr}\{\Pi_{b_*}\varrho \}}-\big(\sum_{b_*}{\rm Re}\big\{{\rm Tr}\{\Pi_{b_*}\Pi_a\varrho\}\big\}\big)^2\Big]^{1/2},
\label{from weak measurement to quantum uncertainty0}
\end{eqnarray}
where $\{\ket{b_*}\}$ is the second basis which achieves the maximum, and we have made use of the Jensen inequality and the completeness relation for the second basis, i.e., $\sum_{b_*}{\rm Tr}\{\Pi_{b_*}\varrho\}=1$, to get the third and fourth lines. Next, applying the Cauchy-Schwartz inequality to the numerator in the first term on the right-hand side, i.e., $({\rm Tr}\{\Pi_{b_*}\Pi_a\varrho \})^2=({\rm Tr}\{(\Pi_{b_*}^{1/2}\Pi_a\varrho ^{1/2})(\varrho ^{1/2}\Pi_{b_*}^{1/2})\})^2\le{\rm Tr}\{\Pi_{b_*}\Pi_a\varrho \Pi_a\}{\rm Tr}\{\varrho \Pi_{b_*}\}$, and using the completeness relation $\sum_{b_*}\Pi_{b_*}=\mathbb{I}$, we finally obtain
\begin{eqnarray}
C_{\rm KD}[\varrho ;\{\Pi_a\}]&\le&\sum_a\big[{\rm Tr}\{\Pi_a^2\varrho \}-{\rm Tr}\{\Pi_a\varrho \}^2]^{1/2}\nonumber\\
&=&\sum_a\Delta_{\Pi_a}[\varrho ], 
\label{weak coherence versus quantum uncertainty}
\end{eqnarray}
where $\Delta^2_{\hat{O}}[\varrho ]:={\rm Tr}\{\hat{O}^2\varrho \}-({\rm Tr}\{\hat{O}\varrho \})^2$ is the quantum variance of $\hat{O}$ in the state $\varrho$.   

We proceed to show that the KD coherence for any quantum state $\varrho$ relative to any reference basis $\{\ket{a}\}$ is always lower than or equal to the $l_1$-norm coherence in $\varrho$ relative to the basis $\{\ket{a}\}$, and they give equal value for $d=2$, i.e., for a single qubit. First, let us consider the general case for $d\ge 2$. From Eqs. (\ref{incompatibility and coherence}) and (\ref{KD coherence}), we have 
\begin{eqnarray}
C_{\rm KD}[\varrho;\{\Pi_a\}]&\le&\max_{\{\ket{b}\}}\sum_a\sum_b\Big|\sum_{a'\neq a}|\varrho_{aa'}||\braket{b|a}| |\braket{a'|b}\Big|\nonumber\\
&=&\sum_{a\neq a'}|\varrho_{aa'}|\max_{\{\ket{b}\}} \sum_b|\braket{b|a}| |\braket{a'|b}|.
\label{step 1}
\end{eqnarray}
On the other hand, using the Cauchy-Schwartz inequality we have $\sum_b|\braket{b|a}| |\braket{a'|b}|\le\big(\sum_b|\braket{b|a}|^2\sum_{b'}|\braket{a'|b'}|^2\big)^{1/2}=1$, where we have made use of the completeness relation for the second basis, $\sum_b\ket{b}\bra{b}=\mathbb{I}$, and the equality is reached when the second basis $\{\ket{b}\}$ and the incoherent basis $\{\ket{a}\}$ satisfies $|\braket{a|b}|=\frac{1}{\sqrt{d}}$ for all $a,b$. Finally, upon inserting into Eq. (\ref{step 1}), we obtain 
\begin{eqnarray}
C_{\rm KD}[\varrho;\{\Pi_a\}]\le\sum_{a\neq a'}|\varrho_{aa'}|=C_{l_1}[\varrho;\{\Pi_a\}],
\label{Theorem 1}
\end{eqnarray}
as claimed. Hence, a non-vanishing KD coherence can be used to detect the $l_1$-norm quantum coherence. Moreover, since a vanishing KD coherence leads to a vanishing $l_1$-norm coherence (property (i)), it is a faithful detector. 

Let us show that the inequality of Eq. (\ref{Theorem 1}) is always saturated for a single qubit, i.e., two dimensional quantum system, with an arbitrary quantum state. Assume first for simplicity that the quantum state of the qubit is pure so that it can in general be written as  
\begin{eqnarray}
\ket{\psi }=\psi_0\ket{0}+\psi_1\ket{1}=\cos\frac{\theta}{2}\ket{0}+\sin\frac{\theta}{2}e^{i\eta}\ket{1},
\label{single qubit state} 
\end{eqnarray}
where $0\le\theta\le\pi$ is the polar angle of the Bloch sphere, $0\le\eta\le 2\pi$ is the azimuthal angle, and $\{\ket{0},\ket{1}\}$ are the eigenstates of the Pauli matrix $\sigma_z$. The $l_1$-norm coherence of the quantum state $\ket{\psi }$ with respect to the incoherent basis $\{\ket{a_z}\}=\{\ket{0},\ket{1}\}$ is thus given by $C_{l_1}[\ket{\psi }\bra{\psi};\{\Pi_{a_z}\}]=2\big|\psi_0\psi_1^*\big|=|\sin\theta|$.

Next, for the purpose of computation of the KD coherence defined in Eq. (\ref{KD coherence}), we express the second basis for the two dimensional Hilbert space $\{\ket{b}\}=\{\ket{b+},\ket{b-}\}$ as: 
\begin{eqnarray}
\ket{b(\alpha,\beta)+}&:=&\cos\frac{\alpha}{2}\ket{0}+\sin\frac{\alpha}{2}e^{i\beta}\ket{1};\nonumber\\
\ket{b(\alpha,\beta)-}&:=&\sin\frac{\alpha}{2}\ket{0}-\cos\frac{\alpha}{2}e^{i\beta}\ket{1}, 
\label{complete set of basis in the x-y plane}
\end{eqnarray}
$0\le\alpha\le\pi$, $0\le\beta\le 2\pi$. We note that upon varying the angles $\alpha$ and $\beta$ over the whole ranges of their values, one scans over all the possible orthonormal bases of the two dimensional Hilbert space. Using this parameterization for the second basis, the KD coherence relative to the basis $\{\ket{a_z}\}=\{\ket{0},\ket{1}\}$ can then be computed straightforwardly to give
\begin{eqnarray}
&&C_{\rm KD}[\ket{\psi }\bra{\psi};\{\Pi_{a_z}\}]\nonumber\\
&=&\max_{\ket{b(\alpha,\beta)}}\sum_{a_z}\sum_b\big|{\rm Im}\{\braket{b|a_z}\braket{a_z|\psi }\braket{\psi |b}\}\big|\nonumber\\
&=&\max_{\alpha,\beta}|\sin\theta\sin(\beta-\eta)\sin\alpha|\nonumber\\
&=&|\sin\theta|=C_{l_1}[\ket{\psi }\bra{\psi};\{\Pi_{a_z}\}].
\label{qubit weak coherence and l1-norm coherence in z basis}
\end{eqnarray} 
Hence, for two dimensional pure state, the KD coherence relative to the incoherent basis $\{\ket{a_z}\}$ is indeed equal to the $l_1$-norm quantum coherence with respect to the incoherent basis $\{\ket{a_z}\}$. 

Let us discuss the geometrical meaning of the above calculation before generalizing the result to arbitrary two dimensional incoherent basis and arbitrary mixed state. First, note that the maximization over the two parameters $\alpha,\beta$ characterizing the second basis $\{\ket{b(\alpha,\beta)}\}=\{\ket{b(\alpha,\beta)+},\ket{b(\alpha,\beta)-}\}$ are carried out independently of each other. The maximization over $\alpha$, which parameterizes the amplitude of $\braket{a_z|b(\alpha,\beta)\pm}$, is obtained for $\alpha=\pi/2$. This means that the basis $\{\ket{b(\alpha,\beta)}\}$ must lie on the the equator of the Bloch sphere so that it is mutually unbiased with the incoherent basis $\{\ket{a_z}\}=\{\ket{0},\ket{1}\}$. Next, the maximization over $\beta$, which parameterizes the phase of $\braket{a_z|b(\alpha,\beta)\pm}$, is obtained for $\beta=\eta+\pi/2$. Combined together, the maximum is attained when the second basis is given by $\{\ket{b_*}_z\}=\{\ket{b_*+}_z,\ket{b_*-}_z\}$, where $\ket{b_*\pm}_z=\frac{1}{\sqrt{2}}(\ket{0}\pm ie^{i\eta}\ket{1})$. Hence, the maximal basis $\{\ket{b_*}_z\}$ is orthogonal to the plane on which both the incoherent basis and the quantum state are lying. One thus finds that the maximal basis $\{\ket{b_*}_z\}$ turns out to be also mutually unbiased with $\{\ket{\psi },\ket{\psi }^{\perp}\}$, where $\ket{\psi }^{\perp}=\sin\frac{\theta}{2}\ket{0}-\cos\frac{\theta}{2}e^{i\eta}\ket{1}$ is the orthonormal pair of $\ket{\psi }$. Moreover, note that the state $\ket{\psi}$ reaches its maximal coherence relative to the basis $\{\ket{a_z}\}$ when $\theta=\pi/2$ so that it is mutually unbiased with both $\{\ket{a_z}\}$ and $\{\ket{b_*}_z\}$. Hence, in this case, the state, the incoherent basis, and the maximal second basis, comprise the three mutually unbiased basis for the two dimensional Hilbert space.    

The computation of KD coherence in Eq. (\ref{qubit weak coherence and l1-norm coherence in z basis}) suggests the following generalization for the pure state of a single qubit relative to any arbitrary incoherent basis. Consider the quantum coherence in the state $\ket{\psi}$ with respect to the incoherent orthonormal basis $\{\ket{a_{\vec{n}}}\}=\{\ket{\vec{n}+},\ket{\vec{n}-}\}$, the complete set of eigenbasis of the Pauli operator $\sigma_{\vec{n}}$ along an arbitrary unit vector $\vec{n}$, i.e., $\sigma_{\vec{n}}=\vec{n}\cdot\vec{\sigma}$, where $\vec{\sigma}=(\sigma_x,\sigma_y,\sigma_z)$. We first express the state as 
\begin{eqnarray}
\ket{\psi}=\psi_{\vec{n}+}\ket{\vec{n}+}+\psi_{\vec{n}-}\ket{\vec{n}-}, 
\label{pure qubit state in general basis}
\end{eqnarray}
where $\psi_{\vec{n}\pm}=\braket{\vec{n}\pm|\psi}$, so that the $l_1$-norm quantum coherence reads $C_{l_1}[\ket{\psi }\bra{\psi};\{\Pi_{a_{\vec{n}}}\}]=2|\psi_{\vec{n}+}\psi_{\vec{n}-}^*|$, where $\Pi_{a_{\vec{n}}}=\ket{a_{\vec{n}}}\bra{a_{\vec{n}}}$. Let us show that this is equal to the KD coherence $C_{\rm KD}[\ket{\psi }\bra{\psi};\{\Pi_{a_{\vec{n}}}\}]$. To do this, we shall use the property (iii), namely 
\begin{eqnarray}
C_{\rm KD}[\ket{\psi }\bra{\psi};\{\Pi_{a_{\vec{n}}}\}]=C_{\rm KD}[U\ket{\psi}\bra{\psi}U^{\dagger};\{U\Pi_{a_{\vec{n}}}U^{\dagger}\}],
\label{property 3 in action}
\end{eqnarray}
where $U$ is an arbitrary unitary operator. Let us further choose a unitary operator: $U=\ket{0}\bra{\vec{n}+}+\ket{1}\bra{\vec{n}-}$, so that we have the following transformation of bases: $U\ket{\vec{n}+}\bra{\vec{n}+}U^{\dagger}=\ket{0}\bra{0}$ and $U\ket{\vec{n}-}\bra{\vec{n}-}U^{\dagger}=\ket{1}\bra{1}$, and the quantum state of Eq. (\ref{pure qubit state in general basis}) is transformed into 
\begin{eqnarray}
\ket{\psi'}= U\ket{\psi}=\psi_{\vec{n}+}\ket{0}+\psi_{\vec{n}-}\ket{1}. 
\label{transformed quantum state of the of qubit}
\end{eqnarray} 
Taking all these into account, Eq. (\ref{property 3 in action}) thus becomes  
\begin{eqnarray}
C_{\rm KD}[\ket{\psi }\bra{\psi};\{\Pi_{a_{\vec{n}}}\}]&=&C_{\rm KD}[\ket{\psi'}\bra{\psi'};\{\Pi_{a_z}\}]\nonumber\\
&=&2|\psi_{\vec{n}+}\psi_{\vec{n}-}^*|=C_{l_1}[\ket{\psi}\bra{\psi};\{\Pi_{a_{\vec{n}}}\}], 
\label{imaginary part of weak value of sigma n and quantum coherence in n basis}
\end{eqnarray}
as claimed. Here, in the second line we have used the previous result for the KD coherence relative to the basis $\{a_z\}=\{\ket{0},\ket{1}\}$, noting Eq. (\ref{transformed quantum state of the of qubit}). Recalling the proof of property (iii) given in Eq. (\ref{proof of the unitary covariant property}), the maximum is obtained when the second basis $\{\ket{b_*}_{\vec{n}}\}$ is given by $\ket{b_*+}_{\vec{n}}= U^{\dagger}\ket{b_*+}_z=\frac{1}{\sqrt{2}}(\ket{\vec{n}+}+ie^{i\eta}\ket{\vec{n}-})$ and $\ket{b_*-}_{\vec{n}}= U^{\dagger}\ket{b_*-}_z=\frac{1}{\sqrt{2}}(\ket{\vec{n}+}-ie^{i\eta}\ket{\vec{n}-})$, where $\eta$ is the relative phase between $\psi_{\vec{n}+}$ and $\psi_{\vec{n}-}$. 

Finally, one can generalize the above proof for the equality between the KD coherence $C_{\rm KD}[\varrho;\{\Pi_a\}]$ and the $l_1$-norm coherence $C_{l_1}[\varrho;\{\Pi_a\}]$ for general density operator $\varrho$ in two-dimensional Hilbert space relative to an arbitrary reference basis $\{\ket{a}\}$. First, taking $\{\ket{a_z}\}=\{\ket{0},\ket{1}\}$ as the incoherent basis, and using the expression of Eq. (\ref{complete set of basis in the x-y plane}) for the second basis, one straightforwardly gets $C_{\rm KD}[\varrho;\{\Pi_a\}]=2|\varrho_{01}|=C_{l_1}[\varrho;\{\Pi_a\}]$, where $\varrho_{01}=\braket{0|\varrho|1}$, and the maximum is obtained for the basis in Eq. (\ref{complete set of basis in the x-y plane}) with $\alpha=\pi/2$ and $\beta=\pi/2-\varphi_{01}$, $\varphi_{01}=\arg\{\varrho_{01}\}$. Using this result, one can then prove the equality between the KD coherence and the $l_1$-norm coherence for general density operator relative to general incoherent basis $\{\ket{\vec{n}+},\ket{\vec{n}-}\}$, by again using the property (iii) of unitarily covariant and choose the unitary that transforms the incoherent basis $\{\ket{\vec{n}+},\ket{\vec{n}-}\}$ to the computational basis $\{\ket{0},\ket{1}\}$.  Hence, for a single qubit, the KD coherence defined in Eq. (\ref{KD coherence}) shares all the monotonic character of the $l_1$-norm coherence with respect to certain classes of incoherence preserving quantum operations \cite{Streltsov review}. 

We further show that for a single qubit, the inequality of Eq. (\ref{weak coherence versus quantum uncertainty}) is also saturated for all pure states. First, without loosing generality, let us take one of the elements of the incoherent basis as the positive $z$-axis of the Bloch sphere. The incoherent reference basis is thus given by $\{\ket{a_z}\}=\{\ket{0},\ket{1}\}$, the complete set of orthonormal eigenvectors of $\sigma_z$. For our purpose, it is convenient to express the general state of the qubit as $\varrho=\frac{1}{2}(\mathbb{I}+r_x\sigma_x+r_y\sigma_y+r_z\sigma_z)$, where $r^2=r_x^2+r_y^2+r_z^2\le 1$. One then directly has 
\begin{eqnarray}
C_{\rm KD}[\varrho;\{\Pi_{a_z}\}]&=&|r_x-ir_y|=\sqrt{r^2-r_z^2}\nonumber\\
&\le&\sqrt{1-r_z^2}=\sum_{a_z}\Delta_{\hat{\Pi}_{a_z}}[\varrho],
\end{eqnarray}
in accord with the inequality of Eq. (\ref{weak coherence versus quantum uncertainty}). The equality is reached for pure state where $r^2=1$, as claimed. Hence, for a single qubit, KD coherence can indeed be seen as the genuine quantum share of the uncertainty out of the total quantum uncertainty quantified by the quantum standard deviation.  

Next, it is instructive to compare the KD coherence defined in Eq. (\ref{KD coherence}) with a quantity defined as \cite{Alonso KD quasiprobability witnesses quantum scrambling,Drori nonclassicality tighter and noncommutativity} 
\begin{eqnarray}
\mathcal{N}[{\rm Pr}_{\rm KD}(a,b|\varrho)]:=\sum_{a,b}|{\rm Pr}_{\rm KD}(a,b|\varrho)|-1. 
\label{KD nonclassicality}
\end{eqnarray} 
$\mathcal{N}[{\rm Pr}_{\rm KD}(a,b|\varrho)]$ quantifies the KD nonclassicity, i.e., the negativity and the nonreality in the KD quasiprobability ${\rm Pr}_{\rm KD}(a,b|\varrho)$ defined over the bases $\{\ket{a}\}$ and $\{\ket{b}\}$ which has been argued to signify the genuine quantum behaviour in broad quantum phenomena. It has been shown in Ref. \cite{Lostaglio KD quasiprobability and quantum fluctuation} that it possesses certain plausible requirements for the quantifier of KD nonclassicality. One finds in particular that KD nonclassicality of Eq. (\ref{KD nonclassicality}) is nonincreasing under decoherence operation as for the KD coherence. An interesting observation is made in Ref. \cite{Drori nonclassicality tighter and noncommutativity} where the Authors consider a depolarizing model of decoherence to show that nonnegativity of the real part of the KD quasiprobability is not sufficient to guarantee a completely incoherent state.  

Now, let us assume that KD coherence relative to the basis $\{\ket{a}\}$ is vanishing, i.e., $C_{\rm KD}[\varrho;\{\Pi_a\}]=0$. Then, by the property (i) of faithfulness, we have $[\varrho,\Pi_a]=0$ for all $a$. In this case, noting that $\Pi_a^2=\Pi_a$, the KD quasiprobability relative to the basis $\{\ket{a}\}$ and any other basis $\{\ket{b}\}$ can be written as 
\begin{eqnarray}
{\rm Pr}_{\rm KD}(a,b|\varrho)=\braket{b|\Pi_a\varrho|b}={\rm Tr}\Big\{\Pi_b\frac{\Pi_a\varrho\Pi_a}{{\rm Tr}\{\Pi_a\varrho\}}\Big\}{\rm Tr}\{\Pi_a\varrho\}. 
\label{joint probability of successive measurement}
\end{eqnarray}
This is just the joint probability to get outcomes $(a,b)$ in the successive measurement of $\{\Pi_a\}$ followed with the measurement $\{\Pi_b\}$ so that it is always real and nonnegative. Hence, in this case, KD nonclassicality is vanishing, i.e., $\mathcal{N}[{\rm Pr}_{\rm KD}(a,b|\varrho)]=0$. One therefore concludes that a nonvanishing KD nonclassicality, i.e., $\mathcal{N}[{\rm Pr}_{\rm KD}(a,b|\varrho)]>0$ implies a nonvanishing KD coherence, i.e., $C_{\rm KD}[\varrho;\{\Pi_a\}]>0$. By symmetry, the former also implies $C_{\rm KD}[\varrho;\{\Pi_b\}]>0$. The implication of this result is that the presence of negativity in the KD quasiprobability ${\rm Pr}_{\rm KD}(a,b|\varrho)$ even when it is real, is sufficient to guarantee the coherence relative to one of the defining bases, say $\{\ket{a}\}$. This is so because one can always vary the other defining basis $\{\ket{b}\}$, so that the KD quasiprobability becomes nonreal, giving a nonvanishing KD coherence.  

\section{Operational and statistical meaning}

One of the important problems in the quantification of quantum coherence is to find a quantifier whose definition translates directly into a set of laboratory operations, without recoursing to quantum state tomography. Such a set of the laboratory operations is then said to give an operational meaning to the coherence quantifier thus defined. Fortunately, there are several schemes to reconstruct KD quasiprobability without recoursing first to the quantum state tomography as elaborated in Ref. \cite{Lostaglio KD quasiprobability and quantum fluctuation}. Two of them are summarized below, focusing on the relevant imaginary part of the KD quasiprobability: one is based on two successive projective measurements proposed by Johansen \cite{Johansen quantum state from successive projective measurement}, and the other is a direct reconstruction based on weak measurement with postselection \cite{Aharonov weak value,Wiseman weak value,Aharonov-Daniel book} suggested by Lundeen et. al. \cite{Lundeen measurement of KD distribution,Salvail direct measurement KD distribution,Bamber measurement of KD distribution,Thekkadath measurement of density matrix}. These schemes for the reconstruction of the KD quasiprobability lend themselves to the operational interpretation of the KD coherence defined in Eq. (\ref{KD coherence}). 

Let us first discuss the method suggested by Johansen based on two successive projective measurements \cite{Johansen quantum state from successive projective measurement}. This is done by noting that the imaginary part of the KD quasiprobability can be expressed as 
\begin{eqnarray}
&&{\rm Im}\{{\rm Pr}_{\rm KD}(a,b|\varrho)\}={\rm Im}\{{\rm Tr}\{\Pi_b\Pi_a\varrho\}\}\nonumber\\
&=&-{\rm Im}\{{\rm Tr}\{\Pi_a\Pi_b\varrho\}\}=\frac{1}{2}{\rm Tr}\{(\varrho_a-\varrho)\Pi_{b|a}^{\pi/2}\}.
\label{Johansen schema for measurement of KD quasiprobability}
\end{eqnarray} 
Here $\varrho_a=\Pi_a\varrho\Pi_a+(\mathbb{I}-\Pi_a)\varrho(\mathbb{I}-\Pi_a)$ is the state of the system after the binary measurement of $\Pi_a$ without learning the outcomes, where $\mathbb{I}-\Pi_a$ is the complement projector to $\Pi_a$, and $\Pi_{b|a}^{\pi/2}=e^{i\Pi_a\pi/2}\Pi_be^{-i\Pi_a\pi/2}$ is the new second basis after a selective rotation generated by the first basis. We note that while performing the selective rotation to obtain $\Pi_{b|a}^{\pi/2}$ is operationally challenging, it in principle can be done. The KD coherence can thus be expressed as, upon inserting Eq. (\ref{Johansen schema for measurement of KD quasiprobability}) into Eq. (\ref{KD coherence}),
\begin{eqnarray}
&&C_{\rm KD}[\varrho;\{\Pi_a\}]=\frac{1}{2}\max_{\{\ket{b}\}}\sum_{a,b}|{\rm Tr}\{[\varrho-\varrho_a]\Pi_{b|a}^{\pi/2}\}|. 
\end{eqnarray} 
Hence, to observe the KD coherence relative to the basis $\{\ket{a}\}$, we need to measure the expectation values of $\Pi_{b|a}^{\pi/2}$ in the states $\varrho$ and $\varrho_a$, compute the difference, and optimize over all possible choices of $\{\ket{b}\}$. In this scheme, KD coherence therefore admits a statistical interpretation as the maximal state disturbance induced by the measurement  $\{\Pi_a,\mathbb{I}-\Pi_a\}$ as observed in the expectation value of $\{\Pi_{b|a}^{\pi/2}\}$. 

Let us proceed to discuss the direct reconstruction of KD quasiprobability via weak measurement with postselection proposed by Lundeen and co-workers \cite{Lundeen measurement of KD distribution,Salvail direct measurement KD distribution,Bamber measurement of KD distribution,Thekkadath measurement of density matrix}. Consider the weak measurement of a Hermitian observable $A$ without significantly perturbing the preselected state $\varrho$, followed by a postselection on a state $\ket{\phi}$ via a normal (i.e., strong) projective measurement. One then obtains the following weak value \cite{Aharonov weak value,Wiseman weak value,Aharonov-Daniel book}: 
\begin{eqnarray}
A^{\rm w}(\phi|\varrho)=\frac{\braket{\phi|A\varrho|\phi}}{\braket{\phi|\varrho|\phi}}. 
\label{complex weak value}
\end{eqnarray}
Note that the weak value $A^{\rm w}(\phi|\varrho)$ may take real numbers outside of the range of the eigenvalues of $A$, and it can even be complex. Such values are called strange or anomalous weak values. The real and imaginary parts of $A^{\rm w}(\phi|\varrho)$ can be inferred respectively from the average shift of the position and momentum of the pointer of the measuring device \cite{Lundeen complex weak value,Jozsa complex weak value}. Noting this, the imaginary part of the KD quasiprobability of Eq. (\ref{Kirkwood-Dirac quasiprobability}) can therefore be directly observed by first weakly measuring $\Pi_a$ with the preselected state $\varrho$, and then followed by the postselection on $\ket{b}$, infer the imaginary part, and multiplied by the probability of the successful postselection, i.e., 
\begin{eqnarray}
{\rm Im}\{{\rm Pr}_{\rm KD}(a,b|\varrho)\}&=&{\rm Im}\{\frac{\braket{b|\Pi_a\varrho|b}}{\braket{b|\varrho|b}}\}\braket{b|\varrho|b}
={\rm Im}\{\Pi^{\rm w}_a(b|\varrho)\}{\rm Pr}(b|\varrho). 
\label{KD quasiprobability from weak measurement}
\end{eqnarray}
The KD coherence $C_{\rm KD}[\varrho;\{\Pi_a\}]$ of Eq. (\ref{KD coherence}) can thus be obtained by taking the sum of the absolute value of Eq. (\ref{KD quasiprobability from weak measurement}), and maximize over all possible choices of the postselection bases:
\begin{eqnarray}
C_{\rm KD}[\varrho;\{\Pi_a\}]=\max_{\{\ket{b}\}}\sum_a\sum_b\big|{\rm Im}\big\{\Pi^{\rm w}_a(b|\varrho)\big\}\big|{\rm Pr}(b|\varrho).
\label{from weak measurement to KD coherence}
\end{eqnarray} 

The above operational interpretation of the KD coherence in terms of the statistics of weak values suggests the following statistical interpretation inherited from the interpretation of the weak value. First, as argued in \cite{Johansen weak value best estimation,Hall prior information,Hofmann imaginary part of weak value in optimal estimation,Agung general c-valued physical quantities and uncertainty relation}, the imaginary part of the weak value $A^{\rm w}(b|\varrho)$ defined in Eq. (\ref{complex weak value}) can be interpreted as the strength of the error in an optimal estimate of $A$ (or a real-deterministic c-valued quantity associated with $A$ and $\varrho$ \cite{Agung general c-valued physical quantities and uncertainty relation}) based on information about $\{b\}$ obtained from a projective measurement $\{\Pi_b\}$, given prior information about preparation represented by $\varrho$. With this in mind, $C_{\rm KD}[\varrho;\{\Pi_a\}]$ obtained operationally in Eq. (\ref{from weak measurement to KD coherence}) can thus be interpreted as the maximum average absolute error of estimating the incoherent basis $\{\ket{a}\}$, by varying the postselection basis $\{\ket{b}\}$, given a preparation associated with the quantum state $\varrho$.   

Hence, the KD coherence $C_{\rm KD}[\varrho;\{\Pi_a\}]$ devised in this work has transparent meanings in terms of direct laboratory operations. It is clear from the above operational schemes to observe KD coherence that the resource consuming procedure is the maximization over all possible second bases $\{\ket{b}\}$. This classical optimization can be done via variational quantum circuits in a hybrid quantum-classical scheme.  Let us note that, at least for a single qubit (two dimensional system), the method of computing e.g. the $l_1$-norm coherence by first reconstructing the density matrix via the state tomography, is much simpler than the above operational schemes either based on two successive measurements or weak measurement with postselection. We emphasize however that the procedure for the state tomography does not tell us the operational meaning of the $l_1$-norm coherence. By contrast, KD coherence translates directly to a set of laboratory operations, leading to their statistical meaning, which might give insight into its application in quantum information processing. 

Moreover, if one only aims to detect the presence of coherence of an unknown quantum state with respect to an incoherent basis $\{\ket{a}\}$, then one may skip the operationally cumbersome maximization over classical parameters. Namely, it is sufficient to find a second basis $\{\ket{b}\}$ so that the $l_1$-norm of the imaginary part of the KD quasiprobability is nonvanishing, i.e. $\sum_a\sum_b\big|{\rm Im}\{{\rm Pr}_{\rm KD}(a,b|\varrho)\}\big|>0$, which, by definition of Eq. (\ref{KD coherence}), guarantees a nonvanishing of the KD coherence, and thus, by virtue of Eq. (\ref{Theorem 1}) guarantees a nonvanishing $l_1$-norm quantum coherence. Since $|{\rm Im}\{{\rm Pr}_{\rm KD}(a,b|\varrho)\}\big|=|{\rm Im}\{{\rm Pr}_{\rm KD}(b,a|\varrho)\}\big|$, it also indicates the coherence with respect to the basis $\{\ket{b}\}$. The maximization over one of the two bases, i.e., over $\{\ket{b}\}$ or $\{\ket{a}\}$, defines the KD coherence with respect to the other basis.    

Having expressed the KD coherence in terms of weak measurement with postselection as discussed above, it still makes sense operationally if the incoherent basis that is given by the set of one-dimensional (rank one) projectors $\{\Pi_{a}\}$ is replaced by a more general measurement basis. This suggests a generalization of the KD coherence as follows. Consider a complete set of POVM measurement, i.e., $\{M_x\}$, $M_x\ge 0$, $\sum_xM_x=\mathbb{I}$. We then define the KD coherence with respect to the POVM basis as 
\begin{eqnarray}
C_{\rm KD}[\varrho;\{M_x\}]&:=&\max_{\{\ket{b}\}}\sum_x\sum_b\big|{\rm Im}\big\{\braket{b|M_x\varrho|b}\big\}\big|\nonumber\\
&=&\max_{\{\ket{b}\}}\sum_x\sum_b\frac{1}{2}\big|\braket{b|[M_x,\varrho]|b}\big|. 
\label{generalized KD coherence}
\end{eqnarray}
Note however that in this case, a state is in general incoherent if $[M_x,\varrho]=0$ for all $x$. $C_{\rm KD}[\varrho;\{M_x\}]$ reduces to Eq. (\ref{KD coherence}) when $\{M_x\}$ is a set of orthonormal one dimensional projectors, but it also covers the case when the rank of the projectors are larger than one, allowing the definition of coherence relative to the decomposition of the Hilbert space into subspaces with dimension larger than one, and also the case when the POVM operators are not orthogonal. See Ref. \cite{Luo-Sun coherence from skew information} for a different approach. Let us for example assume that the POVM is obtained by coarse-graining the incoherent basis, i.e., $M_\mathcal{A}=\sum_{a\in\mathcal{A}}\Pi_a$, where $\mathcal{A}$ is the disjoint subsets partitioning of the indices $\{a\}$. Such a coarse-graining arises naturally if there is a degeneracy. Then, in this case, we have:
\begin{eqnarray}
C_{\rm KD}[\varrho;\{M_{\mathcal A}\}]&=&\max_{\{\ket{b}\}}\sum_{\mathcal{A}}\sum_b\big|{\rm Im}\big\{\braket{b|\sum_{a\in\mathcal{A}}\Pi_a\varrho|b}\big\}\big|\nonumber\\
&\le&\max_{\{\ket{b}\}}\sum_{a,b}\big|{\rm Im}\big\{\braket{b|\Pi_a\varrho|b}\big\}\big|\nonumber\\
&=&C_{\rm KD}[\varrho;\{\Pi_a\}]. 
\label{generalized KD coherence}
\end{eqnarray}
Hence, the KD coherence is nonincreasing under coarse-graining of the incoherent basis. 

As a final note, KD quasiprobability has been argued as a central object in the study of quantum fluctuations arising in a broad field of quantum science \cite{Lostaglio KD quasiprobability and quantum fluctuation}. This observation naturally suggests a possible application of the concept of KD coherence to characterize such quantum fluctuations. Here, we show that it can be used to characterize linear response function. The exposition below follows that of Ref. \cite{Lostaglio KD quasiprobability and quantum fluctuation}. Let us consider a unitary dynamics with the Hamiltonian $H(t)=H_0-\lambda(t) A$, with $A$ a perturbation and $\lambda(t)$ is nonzero only for $t > 0$. Then, in the linear response regime, we have, ${\rm Tr}\{B(t)\varrho(t)\}-{\rm Tr}\{B(0)\varrho(0)\}\approx\int_0^t{\rm d}t'\lambda(t')\Phi_{AB}(t',t)$, where $\varrho(t)$ is the quantum state at time $t$, $\Phi_{AB}(t',t)$ is called linear response function that is given by $\Phi_{AB}(t',t)=i{\rm Tr}\{[A(t'),B(t)]\varrho(0)\}$, with $O(t)=e^{iH_0t}Oe^{-iH_0t}$. Expressing $A(t)=\sum_aa\Pi_{a(t)}$ and $B(t)=\sum_bb\Pi_{b(t)}$, where $\ket{a(t)}=e^{iH_0t}\ket{a}$, $\ket{b(t)}=e^{iH_0t}\ket{b}$, the linear response function can be written in terms of the imaginary part of the KD quasiprobability as 
\begin{eqnarray}
\Phi_{AB}(t',t)=2\sum_{a,b}ab{\rm Im}\{{\rm Pr}_{\rm KD}(a(t'),b(t)|\varrho(0))\}. 
\end{eqnarray}
It encodes the correlation between the observable $B(t)$ and the perturbation. Taking the absolute value, and maximing over all possible $B\in\Lambda_B$ with the same nontrivial spectrum of eigenvalues, one thus obtains 
\begin{eqnarray}
\max_{B\in\Lambda_B}|\Phi_{AB}(t',t)|&\le& 2|a|_*|b|_*\max_{\ket{b(t)}}\sum_{a,b}\big|{\rm Im}\{{\rm Pr}_{\rm KD}(a(t'),b(t)|\varrho(0))\}\big|\nonumber\\
&=& 2|a|_*|b|_* C_{\rm KD}[\varrho(0);\{\Pi_{a(t')}\}], 
\end{eqnarray} 
where $|a|_*$ and $|b|_*$ are the maximum absolute eigenvalues of $A$ and $B$, respectively. Hence, the KD coherence in the initial state relative to the incoherent basis $\{\ket{a(t')}\}$ determines an upper bound to the absolute linear response function maximized over all $B$ with a fixed spectrum. This means that a nonvanishing KD coherence is necessary for a nonvanishing linear response function. 

\section{Summary and Remarks}  

Given a quantum state and an incoherent basis, we have identified a quantity, KD coherence, defined as the $l_1$-norm of the imaginary part of the associated KD quasiprobability defined over the incoherent basis and a second basis, and maximized over all possible choices of the latter. It quantifies the failure of commutativity of the state with the incoherent basis, and satisfies certain desirable properties for a quantifier of coherence. It is upper bounded by the total sum of the quantum standard deviation, i.e., the quantum uncertainty, of the incoherent basis in the state. KD coherence gives a lower bound to the $l_1$-norm quantum coherence, and for arbitrary state of a single qubit, they yield equal values. We demonstrated that KD coherence can be translated directly into laboratory operations, i.e., without recoursing to quantum state tomography, in a couple of quantum-classical hybrid schemes, leading to the statistical meaning as maximum disturbance induced by the measurement of, or as the maximum mean absolute error in the estimation of the incoherent basis. Finally, we discuss the relevance of the KD coherence to characterize the linear response function. We hope our results will initiate a program to use the nonclassicality of KD quasiprobability, and its closely related concept of anomalous weak values, to access various nonclassical aspects encoded in the quantum state such as asymmetry and quantum correlation. It might thus give a better intuition and fresh insight into their roles as resources in quantum information processing, and in wide areas of quantum science where KD quasiprobability has been shown to play important roles \cite{Lostaglio KD quasiprobability and quantum fluctuation}. 

\begin{acknowledgments} 
This work is partly funded by Institute for Research and Community Service, Bandung Institute of Technology, under the program of research assignment with the contract number: 2971/IT1.B07.1/TA.00/2021. It is also in part supported by the Indonesia Ministry of Research, Technology, and Higher Education through PDUPT research scheme with the contract number: 187/E5/PG.02.00.PT/2022 and 2/E1/KP.PTNBH/2019. The Authors would like to thank the anonymous Referees for constructive criticism and suggestions, and Mohammad K. Agusta for useful discussion. 
\end{acknowledgments} 

\appendix

\end{document}